\documentclass[aps,prl,showpacs,preprintnumbers,twocolumn]{revtex4}
\usepackage{epsfig,graphicx}

\input{tcilatex}

\begin{document}

\preprint{CAS-KITPC/ITP-005}

\title{An Ideal Channel of Long Distance Entanglement in Spin Systems}
\author{Xiao-Qiang Xi$^{1,\ 2}$, T. Zhang$^{3}$, R. H. Yue$^{3,\ 4}$, X. C. Xie$^2$, W. M. Liu$^2$}
\address{$^1$Department of Applied Mathematics and Physics,
 Xi'an Institute of Posts and Telecommunications, Xi'an 710061, China}
\address{$^2$Beijing National Laboratory for Condensed Matter Physics,
Institute of Physics, Chinese Academy of Sciences, Beijing 100080,
China}
\address{$^3$Institute of Modern Physics, Northwest
University, Xi'an 710069, China}
\address{$^4$Department of Physics, Ningbo
University, Ningbo 315211, China}

\begin{abstract}
We propose a scheme for using spin chain to realize an ideal channel
of long distance entanglement. The results show that there has
different entanglement in different Hilbert subspace, the
anisotropic parameter $\Delta$ will frustrate the entanglement and
the magnetic field affect the entanglement through changing the
ground state, the boundary entanglement $C_{1N}$ has the simplest
expression in the simplest subspace and it only depend on the first
item of the ground state, that item can be increased when a local
magnetic field is introduced. Our propose can be handled easily
because it only needs a uniform XX open chain initialized in the
simplest Hilbert subspace and a bulk magnetic field that absent for
the boundary qubits.
\end{abstract}

\vspace{0.3cm}  \vspace{0.2cm}

\maketitle

\emph{{Introduction.}} $-$Quantum entanglement has played an
important role in modern physics, for example, it is used to test
some fundamental questions of the quantum mechanics \cite{Bell1987}
and to act a central role in quantum information processing, like
teleportation \cite{Bennett1993}, super-dense
 cording \cite{Bennett1992}, quantum computational speed-ups
\cite{Shor1997, Grover1997}, quantum cryptographic protocols
\cite{Ekert1991,Deutsch1996}, one-way quantum computation
\cite{PRL86_5188} and so on. It is also widely used in sensitive
interferometric measurements
\cite{BotoPRL85_2733,DowlingPRA57_4736,JozsaPRL85_2010,BollingerPRA54R4649}
and studying strongly correlated quantum systems
\cite{PreskillJMO47_127}. Especially, the ground-state entanglement
can be related to quantum phase transition
\cite{OsbornePRA66_032110}, Mott insulator-superfluid transition and
quantum magnet-paramagnet transition. All of the applications of
entanglement are closely dependent on how to produce it. Now, there
are many physical systems suggested to realize entanglement
\cite{Bouwmeester97N390,Turchette98PRL81,Rauschenbeutel00S288,
Yamaguchi99APA68,Makhlin99N398,Sorensen01N409}.

 Spin chain is a
nature candidate for producing pairwise entanglement. It has been
used to quantum information processing
\cite{Loss9899,Imamoglu1999,Wang2001PR,PRL91_207901,PRA69_034304,PRA69_052315}.
For the pairwise entanglement, most of works focused on the
entanglement between the nearest pair, which is basic and will help
us to deeply understand the character of entanglement, and looked
the ways to control and maximize the entanglement, the effectively
controlled factors include temperature, interchange coupling
\cite{Arnesen2000,Wang2001PR}, magnetic field and system impurity
\cite{Wang0105,XiPLA297,AhmadPRA052105}. The ideal entanglement can
be realized for the nearest pair, while the nearest pairwise
entanglement is not enough for the practical applications because
its short distance. Generally speaking, in solid system the
entanglement between a pair decreases quickly as the increase of
distance, so the works about non-nearest pairwise entanglement is
scarce comparing with that of nearest, every non-nearest ideal
entanglement will take great contribution to application of
entanglement. There exist entanglement between the
next-nearest-neighbor qubits in the transverse Ising model, but the
maximal value of entanglement is about $4.3\times 10^{-3}
$\cite{OsbornePRA66_032110}. A scheme is proposed for using a
five-qubit open chain with magnetic and system impurities to realize
a boundary entanglement with maximal value of $\frac{1}{2}$
\cite{ACTA55_3026}. Another scheme is proposed for realizing an
ideal boundary entanglement in four-qubit open chain with symmetry
interaction \cite{CP16_1858}. Venuti {\em et al.}
\cite{PRL96_247206} propose a scheme with a
``strong-weak-strong-weak-\dots'' nearest interaction (dimerization)
and a uniform next nearest interaction to realize a long-distance
entanglement in spin system and further propose it to qubit
teleportation and transfer. The starting point of those works is
analyzing the affect of the interaction to the entanglement, their
different lies in their analytical method.

The aim of this paper is to study how the interaction affect the
non-nearest pairwise entanglement, especially the boundary
entanglement in the open chain. We will analyze the corresponding
results then use them as guidance to construct a practicable ideal
channel of long-distance entanglement.
 The Hamiltonian of Heisenberg XXZ open chain with
impurity is
{\footnotesize
\begin{equation}
H=\sum_{i=1}^{N-1}J_i(\sigma_{i}^{+}\sigma_{i+1}^{-}+\sigma_{i+1}^{+}\sigma_{i}^{-})+
\frac{\Delta}{2}\sum_{i=1}^{N-1}\sigma_{i}^{z}\sigma_{i+1}^{z}
+\sum_{i=1}^{N}B_i\sigma_{i}^{z}.
\end{equation}}
where $J_i$ is the interaction between the $i$-th and $(i+1)$-th
qubits, $\Delta$ is the anisotropic parameter, $B_i$ is the magnetic
field, $\sigma^{\pm}=\frac{1}{2}(\sigma^x\pm i\sigma^y)$,
$\sigma^{x},\sigma^{y},\sigma^{z}$ are the Pauli matrices.

The analytical tool in this paper is the concurrence theory
\cite{PRL80_2245,Hill1997}, $C_{ij}$ denotes the pairwise
entanglement between the $i$-th and $j$-th qubits, which ranges from
0 to 1 is monotonous to the entanglement.

\emph{{Three-qubit Heisenberg XXZ open Chain in the uniform Magnetic
Field.}} $-$When the interaction $J_i$ and the magnetic field $B_i$
are uniform, the eigenvalues of the system are $E_{0,7}=\Delta\mp
3B, E_{1,4}=\pm B, E_{2,5}=-\frac{\Delta_{+}}{2}\pm B,
E_{3,6}=-\frac{\Delta_{-}}{2}\pm B$, where
$\Delta_{\pm}=\Delta\pm\sqrt{8J^2+\Delta^2}$ and the corresponding
eigenvectors are {\footnotesize
\begin{eqnarray}
|\psi_0>&=&|000>,  |\psi_7>=|111>, \nonumber \\
|\psi_1>&=&(-|001>+|100>)/{\sqrt{2}}, \nonumber \\
|\psi_m>&=&c_{m1}(|001>+|100>)+c_{m2}|010> \ (m=2,3),\nonumber \\
|\psi_4>&=&(-|011>+|110>)/{\sqrt{2}}, \nonumber \\
|\psi_n>&=&c_{n1}(|011>+|110>)+c_{n2}|101>\ (n=5,6),
\end{eqnarray}}
where $c_{21}=c_{51}=\frac{\Delta_{-}}{\sqrt{2\Delta_{-}^2+16J^2}}$,
$c_{22}=c_{52}=\frac{4J}{\sqrt{2\Delta_{-}^2+16J^2}}$,
$c_{22}=c_{52}=\frac{4J}{\sqrt{2\Delta_{-}^2+16J^2}}$
$c_{31}=c_{61}=\frac{\Delta_{+}}{\sqrt{2\Delta_{+}^2+16J^2}}$ and
$c_{32}=c_{62}=\frac{4J}{\sqrt{2\Delta_{+}^2+16J^2}}$.

Using the concurrence theory, the entanglement between the boundary
qubits $C_{13}$ can be obtained. The expression of
$C_{13}(B,T,J,\Delta)$ is tedious, the numerical results show that
the anisotropic parameter will frustrate the entanglement, and there
exist a critical magnetic field for $C_{13}$, this phenomena comes
from the change of ground state as the magnetic field variety. When
$\frac{3\Delta+\sqrt{8J^2+\Delta^2}}{4}<B$, the ground state is
$|\psi_0>$, $C_{13}=0$; when
$0<B<\frac{3\Delta+\sqrt{8J^2+\Delta^2}}{4}$, the ground state is
$|\psi_2>$,
$C_{13}=2c_{21}^2=\frac{(\Delta-\sqrt{8J^2+\Delta^2})^2}{8J^2
+(\Delta-\sqrt{8J^2+\Delta^2})^2}$. For more clearly, we plotted the
phase diagram of the ground state and the concurrence $C_{13}(B)$
for some certain $\Delta$:

\begin{center}{\epsfxsize 40mm \epsfysize 32mm \epsffile{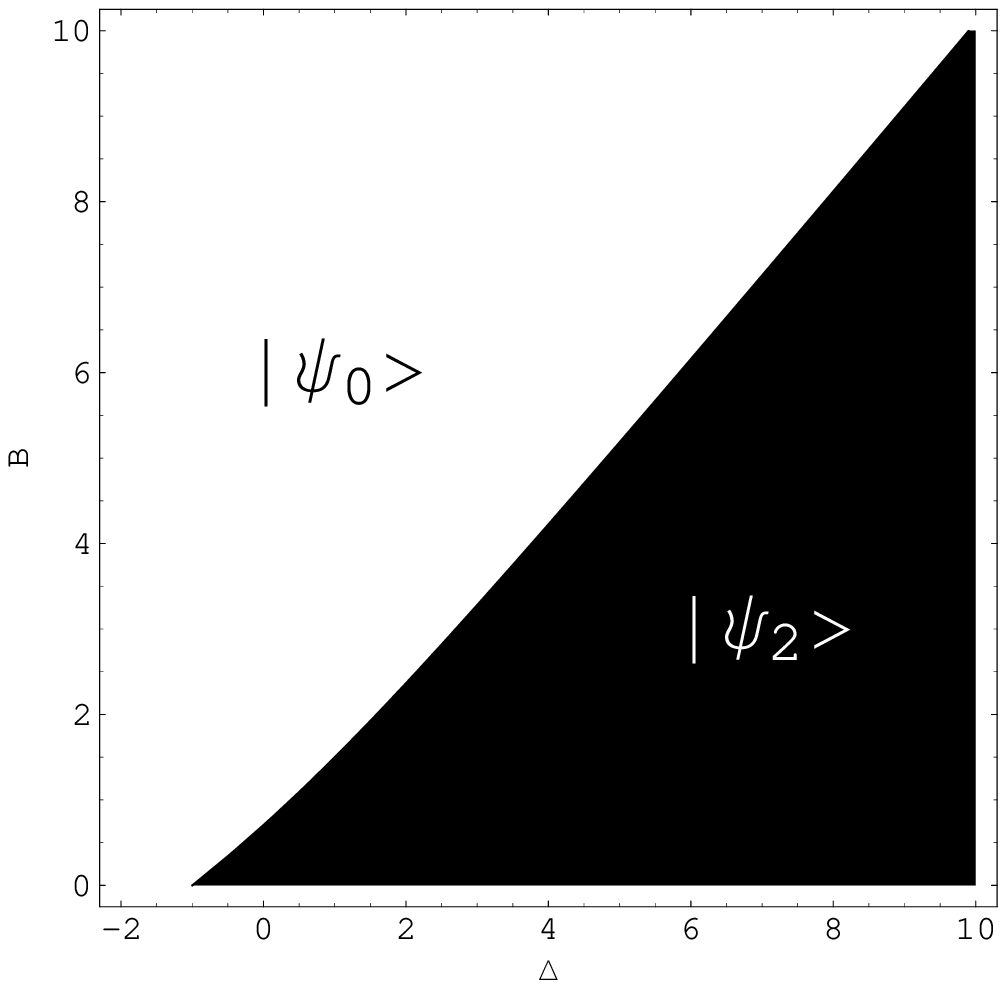}
\epsfxsize 40mm \epsfysize 32mm \epsffile{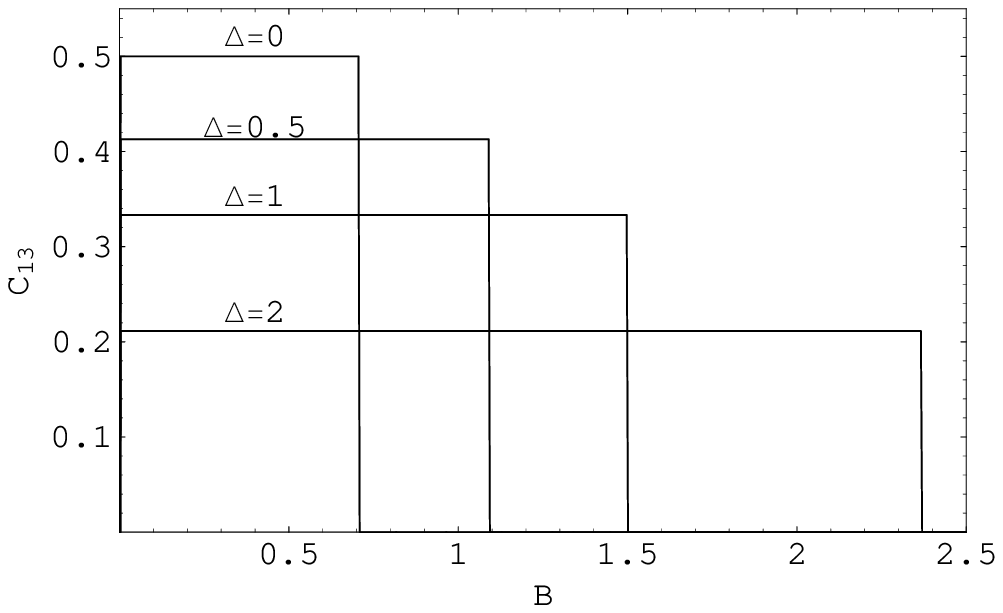}}\\
{\footnotesize(a)\hspace{4cm}(b)} \\
\end{center}
{\footnotesize Figure 1. (a) The phase diagram of the ground state,
the shading is $|\psi_2>$ and the blank is $|\psi_0>$; (b) the
concurrence $C_{13}(B)$ for different $\Delta$.}

When the ground state is $|\psi_2>$, $C_{13}$ decreases as the
increase of $\Delta$, when $\Delta=0,\ C_{13}=\frac{1}{2}$, when
$\Delta=1,\ C_{13}=\frac{1}{3}$, when $\Delta \gg J,\ C_{13}=0$.
i.e. the magnetic field can be used to switch ``on'' and ``off'' the
entanglement. From (b) of Figure 1. we see that the interval of B
(in which the entanglement exist) increases as the increase of
$\Delta$, that is to say, the large interval of entanglement
existing is based on the decrease of entanglement. When $B=0$, the
ground states are $|\psi_2>$ and $|\psi_5>$ (duplicate degeneracy),
$C_{13}=\max\{2c_{21}^2-c_{22}^2,0\}=0$.

\emph{{Four-qubit Heisenberg XXZ Open Chain in the uniform Magnetic
Field.}} $-$As three qubit case, when $J_i$ and $B_i$ are uniform
the eigenvectors can be constructed as {\footnotesize
\begin{eqnarray}
|\psi_0>&=&|0000>,  |\psi_{15}>=|1111>, \nonumber \\
|\psi_m>&=&c_{m1}(|0001>+e^{i\alpha_{m1}}|1000>) \nonumber \\
&& +c_{m2}(|0010>+e^{i\alpha_{m2}}|0100>)\ (m=1,2,3,4), \nonumber \\
|\psi_n>&=&c_{n1}(|1110>+e^{i\alpha_{n1}}|0111>) \nonumber \\
&&+c_{n2}(|1011>+e^{i\alpha_{n2}}|1000>)\ (n=5,6,7,8), \nonumber \\
|\psi_k>&=&c_{k1}(|0011>+e^{i\alpha_{k1}}|1100>) \nonumber \\
&& +c_{k2}(|0101>+e^{i\alpha_{k2}}|1010>) \nonumber \\
&& +c_{k3}(|1001>+e^{i\alpha_{k3}}|0110>) \nonumber \\
&&(k=9,10,11,12,13,14),
\end{eqnarray}}
where the parameters $c_{ij}$, $\alpha_{ij}\ (0,\pi)$ are determined
by $H|\psi>=E|\psi>$ and the normalization condition.

The results of four-qubit case is similar as that of three-qubit,
the magnetic field change the entanglement through changing the
ground state. The entanglement in the ground state is the maximal.
We picked out the possible ground states and their eigenvalues:
$|\psi_{0}>\ (E_0=\frac{3\Delta}{2}-4B)$, $|\psi_{1}>$\ (the state
in $|\psi_m>$ with minimal eigenvalue
$E_1=-\frac{1}{2}(4B+J+\sqrt{5J^2+2J\Delta+\Delta^2})$),
$|\psi_{9}>$\ (the state in $|\psi_k>$ with minimal eigenvalue
$E_{9}$, which can be obtained for concrete $\Delta$ and $J$) and
plotted the phase diagram of the ground state and the concurrence
$C_{14}(B)$ for some certain $\Delta$ in Figure 2.
\begin{center}{\epsfxsize 40mm \epsfysize 32mm \epsffile{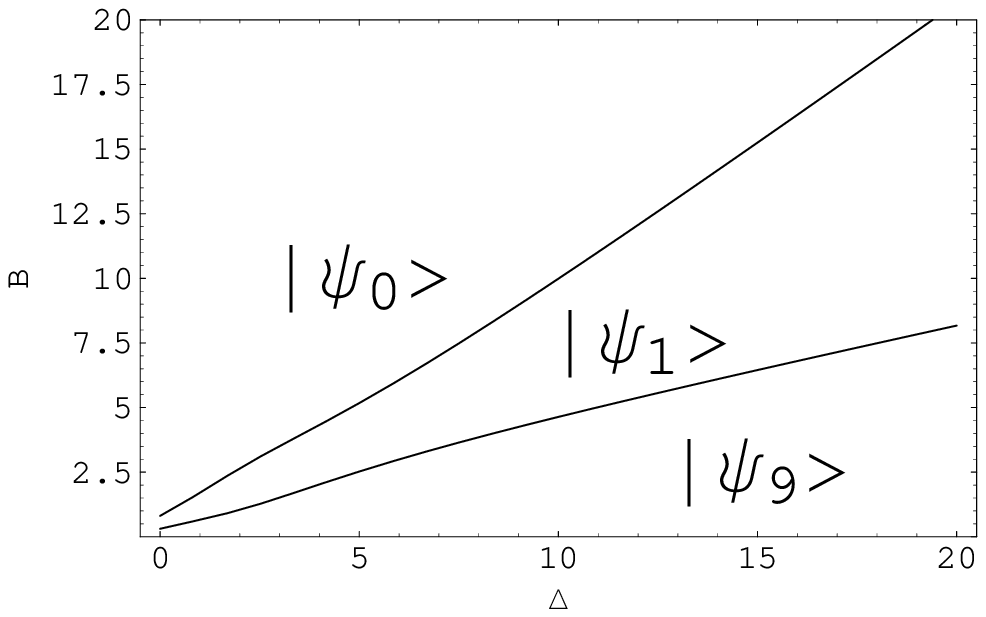}
\epsfxsize 40mm \epsfysize 32mm \epsffile{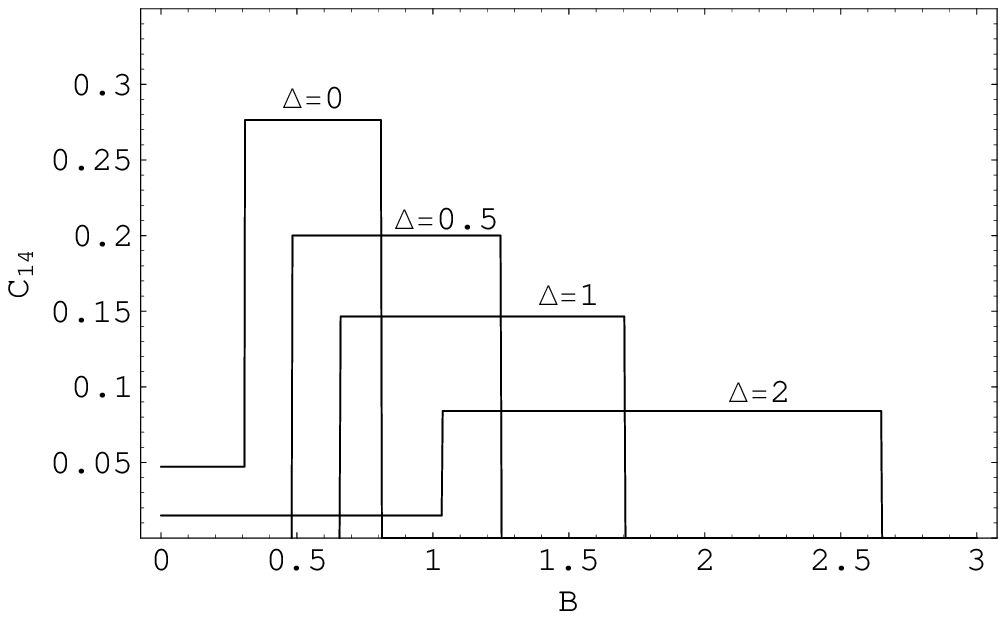}}\\
{\footnotesize(a)\hspace{4cm}(b)} \\
\end{center}
{\footnotesize Figure 2. (a) The phase diagram of the ground state;
(b) the concurrence $C_{14}(B)$ for different $\Delta$.}\\

The maximal value of $C_{14}$ for different $\Delta$ and ground
state are shown in Table 1.

{\footnotesize Table 1. The entanglement $C_{14}$ in the ground
state, ``GS(E)'' instead of the ground state (eigenvalue).}
\begin{center}
{\footnotesize
\begin{tabular}{c|c|c|c}
\hline $\Delta$ & B & GS(E) & ${C_{14}}_{max}$ \cr \hline

0 & $0\leq B<\frac{\sqrt{5}-1}{4}$ & $|\psi_{9}>(-\sqrt{5})$ &
0.0472 \cr

& $\frac{\sqrt{5}-1}{4}<B<\frac{\sqrt{5}+1}{4}$ & $|\psi_{1}>$ &
0.2764 \cr

& $\frac{\sqrt{5}+1}{4}<B$ & $|\psi_{0}>$ & 0 \cr \hline

0.5& $0\leq B<0.48$ & $|\psi_{9}>(-2.712)$ & 0  \cr

& $0.48<B<1.25$ & $|\psi_{1}>$ & 0.2 \cr

& $1.25<B$ & $|\psi_{0}>$ & 0 \cr \hline

 1& $0\leq B<0.66$ &
$|\psi_{9}>(-3.232)$ & 0  \cr

& $0.66<B<1.70$ & $|\psi_{1}>$ & 0.1464 \cr

& $1.70<B$ & $|\psi_{0}>$ & 0 \cr \hline

2& $0\leq B<1.04$ & $|\psi_{9}>(-4.372)$ & 0.0149  \cr

& $1.04<B<2.65$ & $|\psi_{1}>$ & 0.084 \cr

& $2.65<B$ & $|\psi_{0}>$ & 0 \cr \hline
\end{tabular}}
\end{center}
When the ground state is $|\psi_1>$, $C_{14}=2c_{11}^2$ will
decrease as the increase of $\Delta$. When the ground state is
$|\psi_9>$, $C_{14}$ is complex as the variety of $\Delta$:
$C_{14}=0.0472$ if $\Delta=0$; $C_{14}=0$ if $0<\Delta<1$; $C_{14}$
increases as the increase of $\Delta$ if $1<\Delta<7$, $C_{14}$
decreases as the increase of $\Delta$ if $7<\Delta$, the maximal
value is about 0.063.

\emph{{Heisenberg XX Open Chain with system impurity as an ``ideal''
entanglement channel.}} $-$From the conclusion of three- and
four-qubit cases, we see that the boundary entanglement obtain its
maximal value at $\Delta=0$ then decrease as the increase of
$\Delta$. The uniform magnetic field can change the ground state and
the corresponding entanglement. So in this section we no longer
consider the anisotropy parameter $\Delta$ and the magnetic field
$B_i$, XXZ model degenerate into XX model and we also suppose that
all the degeneracy are eliminated.

For convenience, we use total spin to sign the ground state in
different Hilbert subspace, if the ground state has $k$ spin up and
$N-k$ spin down, then total spin
$S_T=\frac{N-k}{2}-\frac{k}{2}=\frac{N-2k}{2}$. For example, the
total spin of $|\psi_0>,\ |\psi_m>,\ |\psi_k>,\ |\psi_n>$ and
$|\psi_{15}>$ in Eq. (3) is $2,1,0,-1$ and $-2$ respectively.
Because the symmetry of spin chain, we only consider the ground
state with total spin
$S_T=\frac{N}{2},\frac{N}{2}-1,-\frac{N}{2}-2,\dots,\frac{N-2[\frac{N}{2}]}{2}$,
the corresponding dimension are $C_N^N,\ C_N^1,\ C_N^2\dots\
C_N^{[\frac{N}{2}]}$ respectively. $S_T=\frac{N}{2}$ is a trivial
subspace with dimension $C_N^N=1$, in the other subspace,
$S_T=\frac{N}{2}-1$ with the minimal dimension $C_N^1=N$ is the
simplest, while $S_T=\frac{N-2[\frac{N}{2}]}{2}$ is the most
complex. The entanglement of the system is bounded by the ground
state entanglement of the subspace.

Without $\Delta$ and $B$, the only interaction left in Eq. (1) is
the exchange hopping $J_i$, i.e.
$H=\sum_{i=1}^{N-1}J_i(\sigma_{i}^{+}\sigma_{i+1}^{-}+\sigma_{i+1}^{+}\sigma_{i}^{-})$.
The case of N=4 is discussed thoroughly in Ref. \cite{Xi0609087}.
Here we study it in the Hilbert subspace. Let $J_1=J_3=1,\ J_2=J$,
for the ground state with $S_T=1$ (i.e. $|\psi_m>$ in Eq. (3), one
spin up),
$C_{14}(1)=2c_{11}^2=\frac{2(J-\sqrt{4+J^2})^2}{2(J-\sqrt{4+J^2})^2+2(2)^2}=\frac{J^2-J\sqrt{4+J^2}+2}
{J^2-J\sqrt{4+J^2}+4}$, ${C_{14}(1)}_{max}=\frac{1}{2}$ when
$J\rightarrow 0$; for the ground state with $S_T=0$ (i.e. $|\psi_k>$
in Eq. (3), two spin up),
$C_{14}(0)=2(|2c_{91}c_{92}|-c_{93}^2)=\frac{J\sqrt{J^2+4}-2}{J^2+4}$,
${C_{14}(0)}_{max}\rightarrow 1$ if $J$ is large enough.

For $N=5$, let $J_1=J_4=1,\ J_2=J_3=J$. Calculations show that
$C_{15}(\frac{3}{2})=\frac{1}{2+4J^2}$ will get its maximal
$\frac{1}{2}$ when $J\rightarrow 0$, the analytical result of
$C_{15}(\frac{1}{2})$ can not be obtained, the numerical results
show that $C_{15}(\frac{1}{2})\rightarrow 1$ when $J\rightarrow
\infty$.

When N=6, let $J_1=J_5=1,\ J_2=J_3=J_4=J$. The analytical result of
$C_{16}$ at the ground state with $S_T=2,1,0$ can not be figured
out. Numerical results show that $C_{16}(2)_{max} \rightarrow
\frac{1}{2}$ when $J\rightarrow 0$; $C_{16}(1)_{max} \rightarrow
0.055$ when $J\approx 2$; $C_{16}(0)_{max}=0.8$ when $J\rightarrow
\infty$. If one let $J_1=J_5=1,\ J_2=J_4=J$ and $J_3=J^2$,
$C_{16}(0)_{max}=1$ when $J$ is large.

Using this method, one can obtain the boundary entanglement of
N-qubit in the case of a matrix with order
$\frac{(N-1)!}{[\frac{N}{2}]!(N-[\frac{N}{2}])!}$ can be
manufactured. In fact only $N\leq 15$ (when $N=15$, the matrix order
is 429, great less than $2^{15}=32768$) case can be calculated.
Although the ideal entanglement spin channel with interaction
$J_i=J_{N-i}=J^{i-1}$ can be realized, this propose has great
disadvantage when N is large, because realizing a very strong
interaction is very difficult. This channel is practicable only for
a short distance entanglement.

\emph{{A really ideal entanglement channel}} $-$A really ideal
entanglement channel must be enough long and easy to manipulate. For
the system with Eq. (1), the affect of $\Delta$, $J_i$ and the
uniform magnetic field to the entanglement has been discussed in the
above sections. They have no direct contribution to realize ideal
entanglement channel. In the Hilbert subspace of N-qubit Heisenberg
XX open, the simplest subspace is $S_T=\frac{N}{2}-1$ except for the
trivial cast $S_T=\frac{N}{2}$, such a subspace can be really
manipulated when N is enough large, so we choose such a subspace as
a candidate for studying the boundary entanglement
($C_{1N}=2c_{m1}^2$), the most important thing is to distinguish the
component $c_{m1}(|1>+e^{i\alpha_{m1}}|N>)$ ($m=1,2,\dots,N$) with
the others, a good idea is to introduce the nonuniform magnetic
field, in which the simplest case is to absent the magnetic field
for the boundary qubits. So the Hamiltonian of the system can be
written as

\begin{equation}
H=J\sum_{i=1}^{N-1}(\sigma_{i}^{+}\sigma_{i+1}^{-}+\sigma_{i+1}^{+}\sigma_{i}^{-})
+B\sum_{i=2}^{N-1}\sigma_{i}^{z}.
\end{equation}

The initial state of the system is prepared as the first qubit to be
spin up and the others spin down. Since the Hamiltonian commutes
with the total spin component along the $z$ direction, the relevant
Hilbert subspace must be spanned by the states
$|j>=|0_1,0_2,\dots,0_{j-1},1_j,0_{j+1},\dots,0_N>$ with
$j=1,\dots,N$. So the eigenvectors of the system can be written as
{\footnotesize
\begin{eqnarray}
|\psi_m>&=&\sum_{j=1}^{k}c_{mj}(|j>+e^{i\alpha_{mj}}|N+1-j>), N=2k
\\
|\psi_m>&=&\sum_{j=1}^{k}c_{mj}(|j>+e^{i\alpha_{mj}}|N+1-j>)
\nonumber
\\
& &+c_{m,k+1}|k+1>, N=2k+1
\end{eqnarray}}
where $c_{mj},\ \alpha_{mj}\ ($ or $\pi), j=1,2,\dots,k,\
m=1,2,\dots,N$, are the parameters determined by $H|\psi>=E|\psi>$
and the normalization condition. Because $N=2k$ is simpler than
$N=2k+1$, so we only consider even N case in the following
discussion.

The maximal entanglement is determined by the ground state, suppose
$|\psi_1>$ is the ground state, then $C_{1N}=2c_{11}^2$ will
approximate to 1 if $\sum_{i=2}^{k}c_{1i}^2\ll c_{11}^2$.

For $N=2k$ case, one need to solve two $k\times k$ matrix, they are
\begin{equation}
M_{\pm}=\left(\begin{array}{ccccccc}
x_1 &J &0&\dots&0&0&0 \\
J&x&J&\dots&0&0&0 \\
0&J&x&\dots&0&0&0 \\
\vdots&\vdots&\vdots&\ddots&\vdots&\vdots&\vdots\\
0&0&0&\dots&x&J&0\\
0&0&0&\dots&J&x&J\\
0&0&0&\dots&0&J&x\pm J\\
 \end{array}\right)_{k\times k}
\end{equation}
with $x_1=-(2k-2)B,\ x=-(2k-4)B$ and find its eigenvalues and
eigenvectors. Only k=2 case can be calculated exactly,
$C_{14}=\frac{(2B-J+\sqrt{4B^2-4BJ+5J^2})^2}{(2B-J+\sqrt{4B^2-4BJ+5J^2})^2+(2J)^2}$,
which is great than 0.99 if $B/J>5$, $C_{14}=0.2764$ if $B=0$. When
$k\geq 3$, one can only obtain the numerical result of $C_{1N}$,
fortunately, the numerical results show that the parameters in the
ground state $|\psi_1>$ satisfy
$\frac{|c_{1i}|}{|c_{1,i+1}|}=\frac{2B}{J}=\beta $,
$C_{1N}=2c_{11}^2=\frac{\beta^{2k}(\beta^2-1)}{\beta^2(\beta^{2k}-1)}$,
which is great than 0.99 if $\beta >10$. Details can be seen from
Figure 3.

\begin{center}{{\epsfxsize 40mm \epsfysize 35mm \epsffile{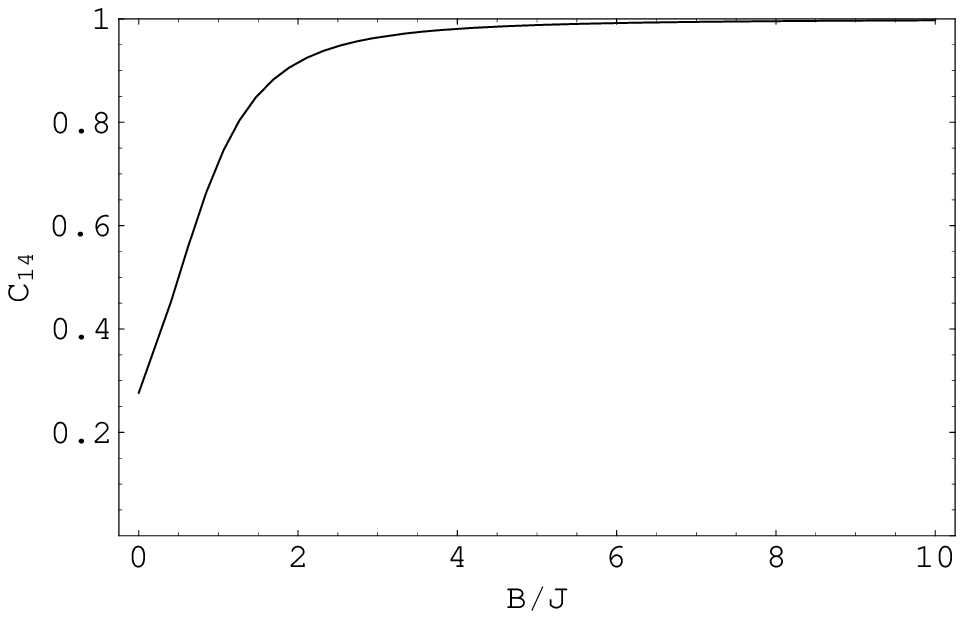}}
\epsfxsize 40mm \epsfysize 35mm \epsffile{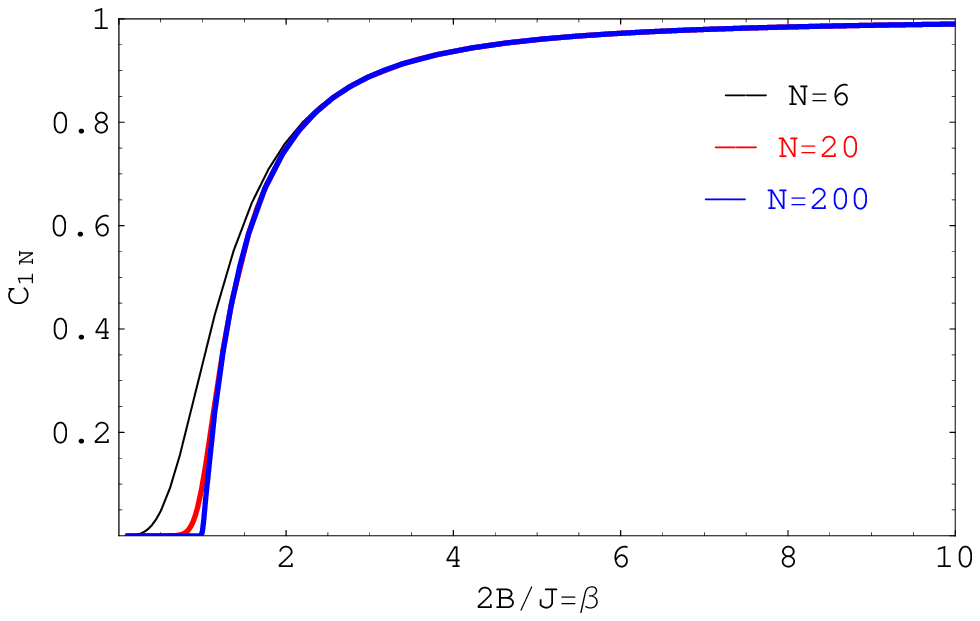}}\\
{\footnotesize(a)\hspace{4cm}(b)}
\end{center}
\begin{center}{\footnotesize Figure 3. (a) The diagram of $C_{14}$
with $B/J$; (b) The diagram of $C_{1N}$ with $2B/J=\beta$.}
\end{center}

In theory, the length of ideal entanglement channel can be infinite
when $2B/J$ is large enough. As the increase of N, the eigenvalue
different between the ground state and excited states will be
smaller and smaller.

\emph{{Conclusions}} $-$The results for three- and four-qubit cases
can tell us the affect of the anisotropic parameter $\Delta$ and
$B_i$ to the entanglement: $\Delta$ will frustrate the boundary
entanglement; the affect of $B_i$ to entanglement is to eliminate
the degeneracy and change the ground state. In fact we can use same
method to calculate N($\leq$15)-qubit cases, the present results are
still valid.

For the Heisenberg XX open chain with nonuniform symmetry
interaction $J_i=J_{N-i}=J^{i-1}$ and $J$ is large enough, the ideal
entanglement can be realized in the most complex subspace, this
conclusion has more theoretical meaning than its application's,
while it is a good candidate if one needs a not too long distance
entanglement.

For the Heisenberg XX open chain with uniform interaction $J$ and a
bulk magnetic field (the boundary qubits are out of the magnetic
field), the long distance ideal entanglement can be realized in the
simplest Hilbert subspace. Our scheme needs two conditions, they are
a uniform XX open with interaction $J$ initialized in the simplest
Hilbert subspace and a bulk magnetic field $B$ absent for the
boundary qubits. Under these conditions $C_{1N}$ will be great than
0.99 if $B/J>5$ for any even N. A uniform interaction chain is
easier to realize than a chain with a
``strong-weak-strong-weak-$\dots-$'' nearest interaction and a
uniform next nearest interaction, in this aspect our scheme is
simpler than that of Venuti {\em et al.}'s \cite{PRL96_247206}. If
our scheme can be realized in experiment, then teleportation in a
solid system will become reality.

This research is supported in part by the Project of Knowledge
Innovation Program (PKIP) of Chinese Academy of Sciences, by the NSF
of China under grant 10547008, 90403019, 90406017, 60525417, by the
National Key Basic Research Special Foundation of China under
2005CB724508, 2006CB921400, by the Foundation of Xi'an Institute of
Posts and Telecommunications under grant 105-0414.

\end{document}